\magnification 1200
\nopagenumbers

\centerline {\bf {Explicit Examples on Conformal Invariance}}
\vskip .7truein
\centerline { M.Horta\c csu * }
\vskip .4truein
\centerline {  Physics Department, Faculty of Sciences and Letters}
\vskip .1truein
\centerline { ITU 80626 Maslak, Istanbul, Turkey }
\vskip 1.3truein
\centerline {\bf { Abstract}}
We study examples where conformal invariance implies rational critical indices, triviality of the underlying quantum field theory and emergence of hypergeometric functions as solutions of the field equations

\vskip .3truein
\noindent
key words: rational critical indices, triviality, conformal symmetry   
\vskip 1.5truein
\noindent
* hortacsu@itu.edu.tr
\baselineskip=18pt
\footline={\centerline {\folio}}
\pageno=1
\vfill\eject

In this note we want to give some examples, obtained from Lagrangian field theory, for results derived by using more formal methods.  These examples, when properly reinterpreted, serve as illustrations for these phenomenae.  

We first want to focus on the  relationship between rational critical indices and the presence of conformal symmetry in a field  theoretical model , a problem which has been studied through several decades.  A classical  paper  on this relation is the one written by Friedan-Qui-Shenker $^{/1}$ where it is shown that the presence of conformal symmetry in statistical mechanical models forces the critical indices for phase transitions to be rational numbers. The conformal symmetry was put to the model through the commutation relations obeyed by the generators, the so called Virasoro algebra and the calculation was dependent on these commutations relations obeyed by the algebra.  The models studied did not have explicit lagrangians.  The correspondence with lagrangian models was through matching dimensions of known exactly solvable  models .   Here unitarity was the forcing constraint resulting in the rational values for the critical indices.  When  c, c being the coefficient of the central term in the Virasoro algebra, is greater than unity, we get field theoretical models instead of the statistical mechanical ones.  In this case the unitarity condition does not impose new constraints; so, one gets a continuous set of solutions. 

Previous calculations by Belavin-Polyakov-Zamolodchikov $^{/2}$ showed the occurrence of rational indices explicitly in their conformal invariant models.  The analysis, again, was not through starting with a lagrangian system.   Both of these papers were working in two dimensional Euclidean spaces. 

It is well known that the conformal group is much richer  in two dimensions compared to higher dimensions.  In two dimensions, it is infinite dimensional.  When the space has a Euclidean metric, any  function that obeys the Cauchy -Riemann equations is locally conformal invariant.  If we use the definition of conformal symmetry as used in general relativity, being the transformation that keeps the light-cone invariant, we see that any metric in two dimensions can be put to this form easily.    We can  find a transformation that sends our former metric to the same metric times a function.   This property is not true for higher dimensional metrics, though. 

A stricker definition of conformal symmetry will be the one defined using  group theory.  This definition coincides with the former one if the space is  higher than two dimensional.  Here we take the conformal group as SO(d,2), where d is the dimension of space-time.  The transformation is expressed as the full Poincar\' e group , scaling transformations where $x^{\mu}$ goes to $\lambda x^{\mu}$, here $\lambda $ is a constant, and the special conformal transformations, where 
$$x^{\mu} \rightarrow {{x^{\mu} - b^{\mu} x^2}\over {1-2bx+b^2x^2}} . \eqno {1}$$
The group has 
$$ {{(d+2)(d+1)}\over {2}} \eqno{2} $$
parameters and $x^2= x_{\mu} x^{\mu} , bx= b_{\mu} x^{\mu} $.  
The special conformal transformations are generated by 
$$ R_{\mu}= P_{\mu}+IP_{\mu}I  \eqno {3}$$ where $P_{\mu}$ is the generator of translations and I takes $x$ to ${{-x}\over{x^2}} $ .  The inversion is not a part of the connected part of the conformal group , except for free theories.

It will be of interest if the relationship between the rationality of the critical indices and the presence of conformal symmetry persists in higher dimensions as well. Schroer and Swieca studied the presence of global conformal symmetry in field theory in 1974 $^{/3}$ .  Using their result, we see that the conformal blocks of Belavin-Polyakov-Zamolodchikov are not local fields, but local fields can be constructed out of them by summing over blocks$^{/4}$ .  It was shown earlier $^{/5}$,  that global conformal invariance is absent in a local field theoretical model unless the two-point function can be written as $(x-x')^{-2n}$  where n is a positive integer.  In this paper two exactly solvable interacting models, as well as  free fields were studied. The essence of the argument is the fact that $R_{\mu}$ generates a kind of conformal "time"  rotations $^{/4}$, and maps spacelike distances into timelike distances.  Unless the two point function has support only on the light-cone, which is the case only when n is an integer,  the special  conformal transformation violates causality and the Wightman functions, therefore the whole theory is not invariant under conformal transformations. When n is an integer, we get a theory which is unitarily equivalent to a theory which is made out of product of free fields $^{/6}$.    

Back in 1970's  , G\" ursey  and Orfanidis $^{/7}$ gave all the representations of the group SO(2,2) , the conformal group in two dimensions in the stricker sense.  They showed that  global conformal invariance exists only when one uses the analytic series out of all possible ones.   The analytic series for SO(2,2) were related to non  trivial interacting field theories with  two-point functions with integer powers, giving explicit examples of the presence of global conformal invariance in two models $^{/8}$, the Thirring and the derivative coupling models.  It was shown that    the representations of SO(2,2) allowing such a symmetry correspond to discrete values of the coupling constant and spin, those values that will give two point functions with integers powers. The Klaiber solution $^{/9}$ to the Thirring model uses two constants  that are functions of the coupling constant "g" and spin "s".  Since the model is in two dimensions, spin does not have a physical meaning and , in principle, can take any value. When spin is equal to one half, as is the case for  any true "spinor",  it is not possible to satisfy conformal invariance.  Only for anomalous values of the spin value,   conformal invariance can exist for certain discrete values of the coupling constant.   

Here what is meant by " non trivial field theories" is just a theory whose coupling constant is not zero.  We know that this is not a sufficient condition for a model to be "nontrivial" $^{/10}$.  There were attempts to carry out calculations for the electron-positron annihilation and electro-production processes, using Thirring model as the fundamental model describing these interactions $^{/11}$.  The results were inconclusive, though. In the former case the cross-section was found to be proportional to a power of the transferred momenta.  In the latter case the presence of generalized hypergeometric functions $^{/12}$ made a clear interpretation not possible.  Thus we could not check whether these processes, based on this conformal invariant model, gave results different from the free field case.

Schroer has shown recently $^{/13}$    that conformal invariance, realized in Minkowski space, does not allow a particle interpretation unless one has a free field theory.  One is led to a trivial theory by the vanishing of the LSZ limit $^{/14}$.  Schroer also comments on $^{/13}$ how the presence of conformal symmetry maps the long and short distance behaviour into each other, resulting in the coalescing of these these two points.  All multi-particle thresholds collapse on top of each other resulting in anomalous dimensions.  In the presence of anomalous dimensions, the LSZ limit vanishes, resulting in the loss of particle interpretation.

Examples of how conformal symmetry leads to a trivial theory were given earlier.   There were attempts to regularize conformal invariant models in d=4 by unconventional models $^{/15}$.   These models were shown to give rise to trivial theories $^{/16}$ in the sense that physical processes, calculated using such models, gave the free field result. In this reference, two physical processes, electron-positron annihilation and the quark-electron structure functions were calculated using the 1/N expansion .  Calculations were carried up to three loops, consistent with the 1/N expansion.  The end result was exactly the one given by the free quark model.  In the annihilation cross-section, the one loop is the free quark result.  The higher loop contributions    cancelled one another, ending up  with  the free quark result.  In the latter case, only the lowest order tree diagram remained, all the diagrams with loops cancelling each other.  The model scaled exactly in both of the processes studied . The logarithmic corrections of QCD were lacking. In this sense this result confirms Schroer's ideas about the triviality of field theoretical models possessing conformal symmetry.  The presence of conformal symmetry in the model results in a free field theory, at best a generalized free field theory which is formed by the powers  of free fields $^{/6}$.   

Schroer $^{/17}$ thinks that, if one starts with AdS space, instead of the Minkowski, there may be a way out.  The zero component of the generator of special conformal transformations, $R^{0}$, may act as the "Hamiltonian in the new space.  At this point we want to remark on examples using models based on the AdS spaces.

Recently there were papers on models where critical behaviour was studied for the emergence of black holes by the presence of scalar fields.  For amplitudes of scalar fields which are less than a critical value, no black-hole exists.  As the amplitude of the scalar field increases, there is a certain value beyond  which we find a black-hole which swallows the scalar field. Such models show what is called Choptuik scaling $^{/18}$.  During  last year the BTZ black hole $^{/19}$ and the closely related AdS  models were studied in the presence of a scalar field. Two sets of results exist.  When one takes a theory built around either the BTZ black hole $^{/20}$   or the AdS solution $^{/21}$ , one gets a fraction, actually just 1/2 for the critical index for black-hole formation. We know that the d=3 AdS solution is dual to a conformal invariant model in d=2 ( Maldecena conjecture $^{/22}$) .  We also know that the BTZ black hole is related to a conformal invariant model $^{/23}$  .When the same model is studied numerically, though, i.e. when one is not expanding around one of these two limiting cases, which takes the scalar field vanishing, one gets a non rational  value for the same index $^{/24,25}$.  Since the index is calculated using numerical methods, one may not be sure if it can still be written as a rational number.  At least it is certain that it is not equal to 1/2, like the numbers obtained in mean field theory  calculations for critical indices.  Just note that mean field theory calculations are independent of any dimensions.  Mean field theory  always has a Gaussian fixed point which  may be considered as one of the zeroes of the beta function of  the related field theory where one obtains exact conformal symetry. 

In the cases when the model can be related to the conformal symmetric theories, the hypergeometric function, the signature of conformal symmetry emerges as the solution of the problem. We see the same function whether we study fluctuations in the background field of an instanton for the Yang-Mills theory $^{/26}$, or try to solve the Seiberg-Witten   $^{/27}$
relations using differential equations $^{/28}$.  We may also encounter the hypergeometric functions while solving critical indices for BTZ solution $^{/20}$ or for statistical models, not mentioning finding their hyperforms when we have more than one relevant variable $^{/11}$.

As a last example we give the emergence of this function if one expands around the exact solution for a metric in three dimensions.
If one starts from a three dimensional metric of the form $^{/24}$
$$ ds^2= {{e^{2A(r)}}\over {cos^2 (r) }} (dr^2 -
dt^2) + tan^2 (r) e^{2B(r)} d\theta^2 ,\eqno{4}$$
which describes a AdS or BTZ solution depending on the solution one uses for the function  $A$ and $B$.
An exact solution , describing the AdS solution is given by $^{/24,29}$
$$ A= -log (sin (r)),\eqno{5}$$
$$ B= log({{cos (2r}) \over {2sin^2 (r)}}).\eqno{6} $$
 If one expands the functions $A$ and $B$ in a power series $$ A= A^0+\epsilon A^1  + ......   , \eqno{7}$$
$$ B= B^0+\epsilon B^1  + .......  , \eqno{8} $$
using the zeroth order solutions as the one given above, one finds
$$ A^{1}_{,rr} - {8A^{1} \over {sin^2 (2r)}} =0 \eqno{9} $$
which can be reduced to an equation of the hypergeometric type $^{/29}$.  A simple calculation shows that 
the solutions can be written in terms of hypergeometric functions, 
$$ sin^{-1} (2r) \times _2F_1(-1/2,-1/2|-1/2|sin^2 (2r) ) \eqno{10}$$ for one solution, and 
$$ sin^2 (2r) \times _2F_1(1,1 | 5/2| sin^2 (2r)) \eqno {11}$$ 
for the other.
In this example, too, we see the mark of conformal symmetry, the hypergeometric functions.

In this note we gave examples of models exhibiting conformal invariance.  If one signature of this symmetry is rational critical indices, the other one is the emergence of the hypergeometric functions, simple or hyper ones, in the solutions.  As shown in reference 24, when the symmetry is not present, for finite values of the $\phi$ field, the solution is only numerical and the index is not rational.  Similarly, if we do not bring in $r$ dependent regulators, as t'Hooft does $^{/26}$, but insert only mass terms, as it is usually done, one breaks the symmetry and loses the hypergeometric functions for the instanton calculation.

\vskip 1.truein

\noindent
{\bf{Acknowledgement}}
This work is partially supported by TUBA, the Academy of Sciences of Turkey and TUBITAK, the Scientific and Technical Research Council of Turkey. I thank Prof.s A.N.Berker and A.Erzan for  important information on   mean field theories. 
\vfill\eject

REFERENCES

\item {1.}    D.Friedan, Z. Qui and S. Shenker, Physical Review Letters {\bf{52}} (1984) 1575;

\item {2.}  A.A.Belavin, A.M.Polyakov and A.B. Zamolodchikov, Nuclear Physics B{\bf{241}} (1984) 333;

\item {3.}   B.Schroer and J.A. Swieca, Physical Review, {\bf{D10}} (1974) 480, B.Schroer,J.A. Swieca and A.H. Volkel, Physical Review, {\b{D11}} (1975) 11;

\item {4.} B. Schroer, hep-th/0010290;

\item {5.} M.Horta\c csu, B.Schroer, R. Seiler, Physical Review {\bf{D5}}(1972) 2519;  

\item {6.} A.S.Wightman,{\it{Carg\` ese Lectures in Physics, 1964}}, edited by B.Janovici (Gordon and Breach) New York, 1967;

\item {7.} F.G\" ursey and S.Orfanidis, Physical Review {\bf{7}}(1973) 2414;

\item {8.} M.Horta\c csu, Il Nuovo Cimento {\bf{17A}} (1973) 411;

\item {9.} B.Klaiber, {\it{ Lectures in Thoretical Physics}}, edited by W.E.Brittin and A.O.Barut, Vol. {\b{XB}} (New York, 1968);

\item {10.}  G.A.Baker and J.M. Kincaid, Physical Review Letters {\bf{42}} (1979) 1431, J. Statistical Physics, {\bf{24}} (1981) 469;

\item {11.} M.Horta\c csu, Bo\~ gazi\c ci University Journal-Sciences, {\bf{1} }  (1973) 15;

\item {12.} P. Appell, J. Kamp\' e de F\' eriet, {\it{ Fonctions Hypergeometriques et Hypersph\' eriques}}, Paris (1926), P. Humbert, Proc. Roy. Soc. Edinburgh,{\bf{41} }(1920-1) 73;

\item {13.}  B.Schroer, Physics Letters, {\bf{B494} } (2000) 124, also hep-th/0005134;

\item {14.} K. Pohlmeyer, Communications in Mathematical Physics, {\bf{12} }  (1969) 201,D.Bucholz and K.Fredenhagen, Journal of Mathematical Physics, {\bf{18} }   (1977) 1107;

\item {15.}  K.G.Akdeniz, M. Ar\i k, M.Durgut, M.Horta\c csu, N.K. Pak, S.Kaptano\~ glu,Physics Letters {\bf{B116} } (1982) 34, 41, K.G.Akdeniz, M. Ar\i k, M.Horta\c csu, N.K. Pak, Physics Letters {\bf{B124}} (1983) 79, M.Ar\i k, M.Horta\c csu and J.Kalayc\i , J. Physics G: Nuclear Physics, {\bf{11}} (1985) 1;

\item {16.}  M.Ar\i k and M.Horta\c csu, J. Phys. G: Nucl. Phys. {\bf{9}} (1983) L119, M.Horta\c csu, Bulletin of the Technical University of Istanbul, {\b{47}} (1994) 321;

\item {17.} B.Schroer, Physics Letters {\b{9B494}} (2000) 124, hep-th/0005134, also K.H.Rehren, hep-th/9905179, hep-th/0003120;

\item {18.} M.Choptuik, Physical Review Letters {\bf{70}} (1993) 9;

\item {19.} M.B\~ anados, C.Teitelboim and J.Zanelli, Physical Review Letters {\bf{69}}(1992) 1849,  M.B\~ anados, M.Henneaux, C.Teitelboim,J.Zanelli, Physical Review {\bf{D48}} (1993)1506;

\item {20.} D.Birmingham, hep-th/0101194, also D.Birmingham, I.Sachs and S.Sen, hep-th/0102155;

\item {21.} W.T.Kim and J.J.Oh, hep-th/0105112;

\item {22.} J.Maldacena, Adv. Theor. Math. Phys. {\bf{2}} (1998) 231, S.S.Gubser, I.R. Klebanov and A.M.Polyakov, Physics Letters {\bf{B428}} (1998) 105; E.Witten, Adv. Theor. Math. Phys. {\bf{2}} (1998) 253;

\item {23.} G.T.Horowitz and D.L.Welch, Physical Review Letters {\bf{71}} (1993) 328, N.Kaloper, Physical Review {\bf{D48}} (1993) 2598;

\item {24.} F.Pretorius and M.Choptuik, Physical Review {\bf{D62}} (2000) 124012; 

\item {25.} D.Garfinkle, Physical Review {\bf{D63}} (2001) 044007, V.Husain and M.Olivier, gr-qc/0008060, L.M.Burko, Physical Review {\bf{D62}} (2000) 127503;

\item {26.}  Gerard t'Hooft, Physical Review {\bf{D14}} (1976) 3432;

\item {27.} N.Seiberg and E.Witten, Nuclear Physics B{\bf{426}} (1994) 19;

\item {28.} Adel Bilal, hep-th/9601007;

\item {29.} T. Birkandan, M.Horta\c csu, gr-qc/0104096.
\end